%% file: CKM_celis.tex
\documentclass[12pt]{article}
\usepackage{graphicx}
\usepackage{amssymb,amsmath}
\usepackage{mathrsfs}
\usepackage{graphicx,subfigure}

\usepackage{cite}
\usepackage{epsfig}

\newcommand{\Br}{\mathrm{Br}}

\newcommand\pubnumber{IFIC/14-65\\FTUV/14-1007}
\newcommand\pubdate{\today}

\def\napoli{IFIC, Universitat de Val\`encia -- CSIC\\ Apt. Correus 22085, E-46071 Val\`encia, SPAIN}
\def\support{\footnote{Work supported by the Spanish Government and ERDF funds from the EU Commission [Grants FPA2011-23778 and CSD2007-00042 (Consolider Project CPAN)].   }}

\def\Title#1{\begin{center} {\Large #1 } \end{center}}
\def\Author#1{\begin{center}{ \sc #1} \end{center}}
\def\Address#1{\begin{center}{ \it #1} \end{center}}

\newcommand\pubblock{\rightline{\begin{tabular}{l} \pubnumber\\
         \pubdate  \end{tabular}}}
\newenvironment{Abstract}{\begin{quotation}  }{\end{quotation}}
\newenvironment{Presented}{\begin{quotation} \begin{center} 
             PRESENTED AT\end{center}\bigskip 
      \begin{center}\begin{large}}{\end{large}\end{center} \end{quotation}}
\def\Acknowledgements{\bigskip  \bigskip \begin{center} \begin{large}
             \bf ACKNOWLEDGEMENTS \end{large}\end{center}}

\input econfmacros.tex

\begin{document}
\begin{titlepage}
\pubblock

\vfill
\Title{$B \rightarrow D^{(*)} \tau \nu$ decays in the aligned two-Higgs-doublet model}
\vfill
\Author{ Alejandro Celis\support}
\Address{\napoli}
\vfill
\begin{Abstract}
In this talk I review the status of $B\rightarrow D^{(*)} \tau \nu$ decays within the framework of the aligned two-Higgs-doublet model.    
\end{Abstract}
\vfill
\begin{Presented}
The 8th International Workshop on the CKM Unitarity Triangle (CKM 2014)\\
Vienna, Austria, September 8-12, 2014
\end{Presented}
\vfill
\end{titlepage}
\def\thefootnote{\fnsymbol{footnote}}
\setcounter{footnote}{0}

\section{Introduction}
The BaBar collaboration has reported an excess with respect to the Standard Model (SM) in exclusive semileptonic transitions of the type $b \rightarrow c \tau^- \bar \nu_{\tau}$~\cite{Lees:2012xj}.  More specifically, they have measured the ratios
\begin{align}  
R(D) \equiv \dfrac{   \mathrm{Br}(   \bar B \rightarrow D \tau^- \bar \nu_{\tau} ) }{   \mathrm{Br(  \bar B \rightarrow D \ell^-  \bar \nu_{\ell} )} }  \stackrel{\rm BaBar}{=}\;  0.440 \pm 0.058 \pm 0.042 \stackrel{\rm avg.}{=}\; 0.438 \pm 0.056  \,, \nonumber \\
R(D^*) \equiv \dfrac{ \mathrm{Br}(\bar B \rightarrow D^* \tau^- \bar \nu_{\tau})  }{  \mathrm{Br}(\bar B \rightarrow D^* \ell^- \bar \nu_{\ell}) }  \stackrel{\rm BaBar}{=}\; 0.332 \pm 0.024 \pm 0.018  \stackrel{\rm avg.}{=}\; 0.354 \pm 0.026 \,,
\end{align}
normalized by the corresponding light lepton modes~$\ell = e, \mu$.    The second values are the averages with previous measurements by the Belle collaboration~\cite{Adachi:2009qg,Bozek:2010xy}.   The BaBar measurements show an excess of $2.0 \sigma$ ($R(D)$) and $2.7 \sigma$ ($R(D^*)$) with respect to the SM~\cite{Lees:2012xj}.   We consider here the possibility that the observed excess in $R(D^{(*)})$ is due to a charged Higgs contribution entering at tree level.   The analysis presented is done within the framework of the aligned two-Higgs-doublet model (A2HDM)~\cite{Pich:2009sp}, see Refs.~\cite{Jung:2010ik,Celis:2012dk} for details.    Other attempts to explain the excess in these observables using different models can be found for example in Refs.~\cite{Fajfer:2012jt,Datta:2012qk,Crivellin:2012ye,He:2012zp,Tanaka:2012nw,Ko:2012sv,Biancofiore:2013ki,Dorsner:2013tla,Sakaki:2013bfa,Abada:2013aba,Duraisamy:2014sna}.

\section{$B \rightarrow D^{(*)}\tau\nu$ decays in the A2HDM}

The full set of experimental observables considered in our analysis and their respective SM predictions is given in Table~\ref{tab::SMvsEXP}.   We only consider processes mediated at tree-level by the charged Higgs, loop-mediated processes have in general a higher UV sensitivity.    It is worth pointing out:
\begin{itemize}
\item The analysis presented here does not include the latest measurement of $\mathrm{Br}(B^+\rightarrow \tau^+ \nu_{\tau})$ with the semileptonic tagging method using the full Belle data sample~\cite{Abdesselam:2014hkd}.
\item Our SM prediction for $R(D)$ agrees with that in Ref.~\cite{Kamenik:2008tj}.    More recent estimations of $R(D)$ have reduced the discrepancy in this observable to about $1 \sigma$~\cite{Bailey:2012jg,Becirevic:2012jf}.  
\end{itemize}
The inclusion of these points would not make a qualitative difference in our analysis of the A2HDM since $R(D^*)$ is the problematic observable at the moment.   

\begin{figure}[htb]
\centering
\includegraphics[height=2.6in]{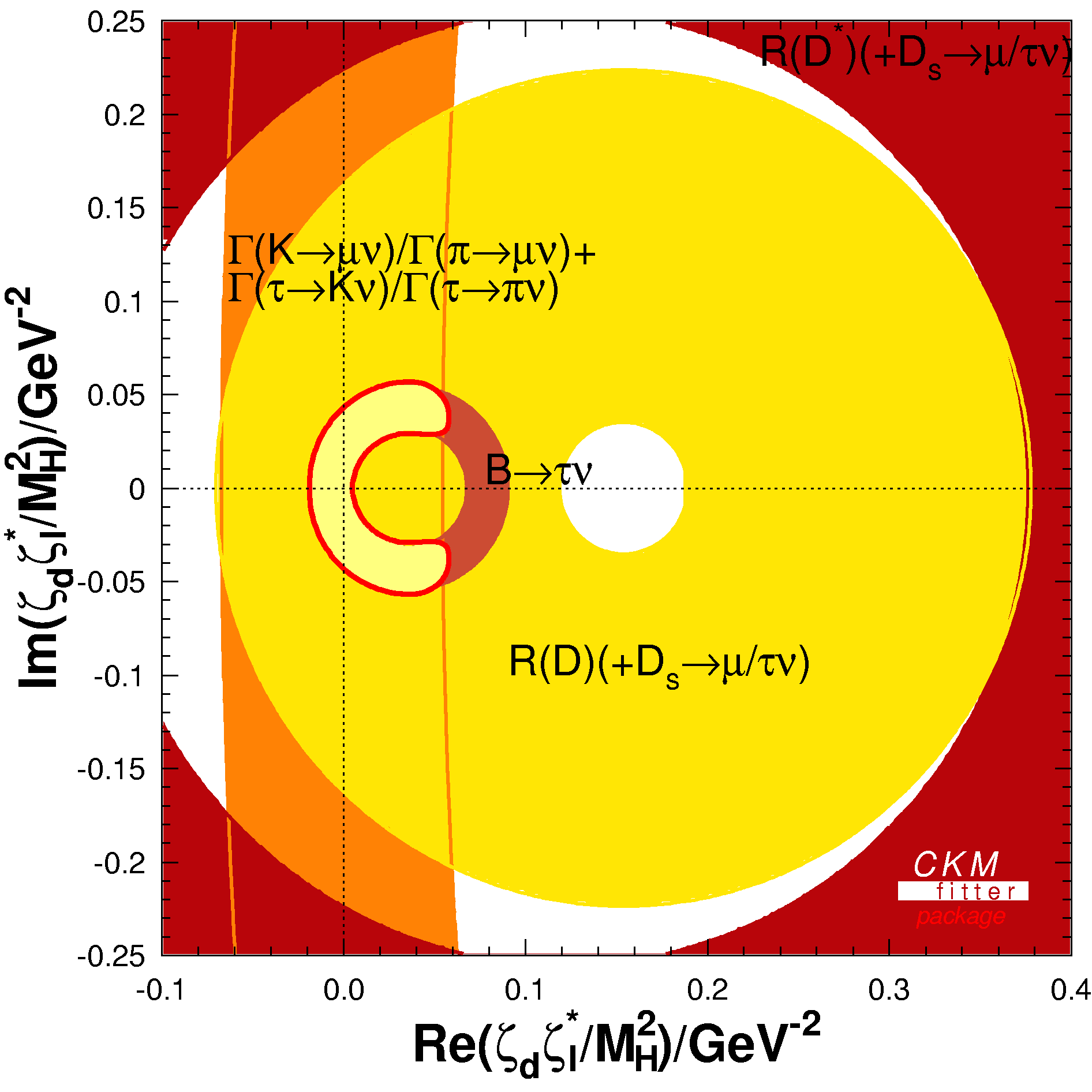}
~
\includegraphics[height=2.6in]{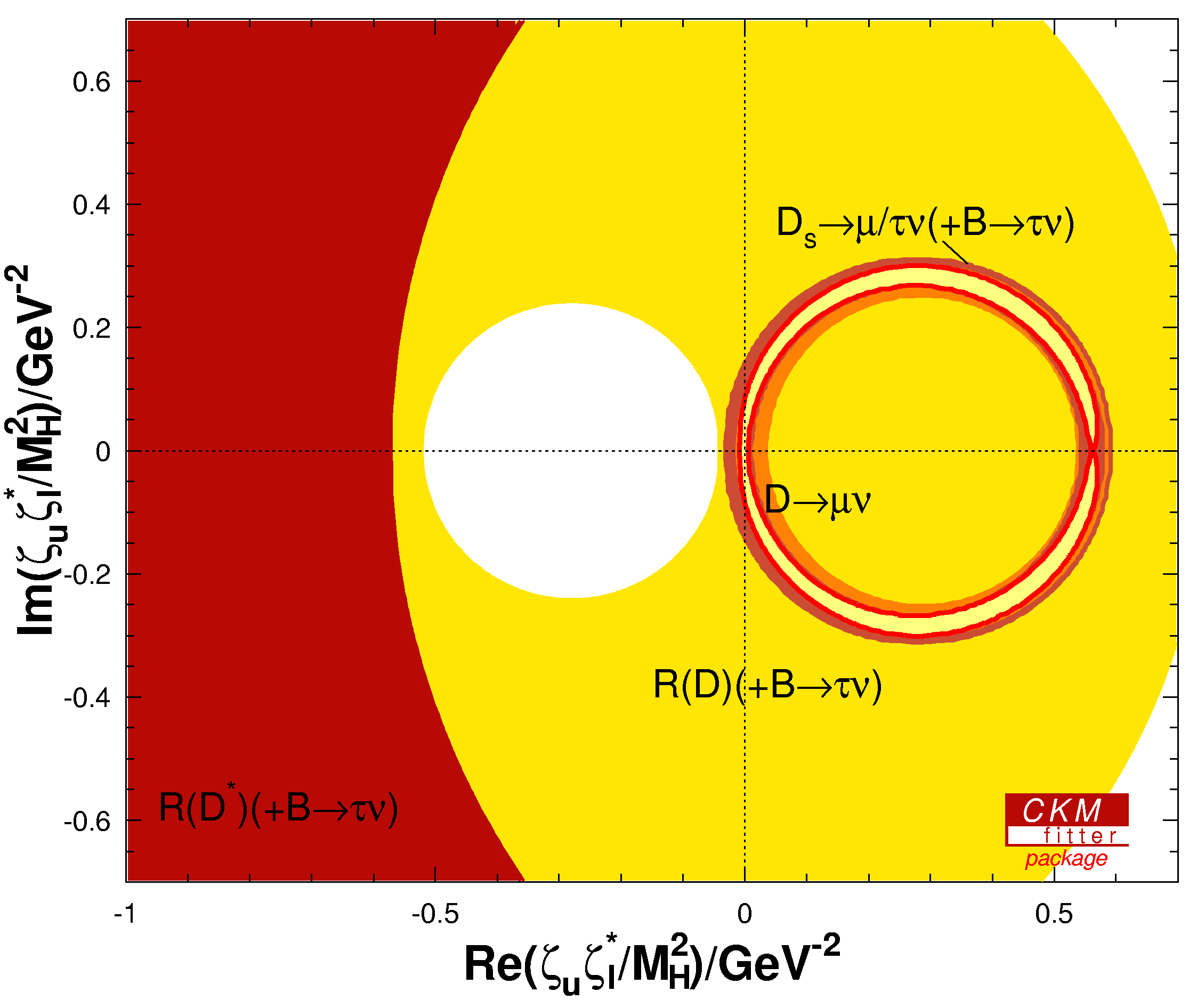}
\caption{\it \small $95\%$~CL allowed regions in the parameter space of the A2HDM by the different observables considered.}
\label{fig:alignment}
\end{figure}

Charged Higgs interactions with fermions are parametrized in the A2HDM by~\cite{Pich:2009sp}
\begin{align}\label{lagrangian}
 \mathcal{L}_Y  =  - \dfrac{\sqrt{2}}{v}\; H^+ \Bigl\{ \bar{u} \left[ \varsigma_d\, V_{\mbox{\scriptsize{CKM}}} M_d \mathcal P_R - \varsigma_u\, M_u V_{\mbox{\scriptsize{CKM}}} \mathcal P_L \right]  d\, + \, \varsigma_l\, \bar{\nu} M_l \mathcal P_R l \Bigr\}
\;  + \;\mathrm{h.c.} \, .
\end{align}
Here $v \simeq (\sqrt{2} G_F)^{-1/2} \simeq 246$~GeV, $M_{u,d,l}$ are the diagonal fermion mass matrices while $V_{\mbox{\scriptsize{CKM}}}$ is the CKM matrix.  Chiral projectors $\mathcal P_{L,R}=(1 \mp \gamma_5)/2$ are denoted as usual.     The family universal alignment parameters $\varsigma_{f}$ ($f=u,d,l$) are independent complex quantities in general.   The different versions of the 2HDM with natural flavour conservation are recovered in specific limits of the A2HDM~\cite{Pich:2009sp}.      The $95\%$~CL allowed regions by the different observables are shown in Figure~\ref{fig:alignment}.   The constraints are shown in the complex planes $\varsigma_d \varsigma_l^*/M_{H^{\pm}}^2$ and $\varsigma_u \varsigma_l^*/M_{H^{\pm}}^2$.     We observe that $R(D^*)+B\rightarrow \tau \nu$ prefer large and negative values for $\mathrm{Re}(\varsigma_u \varsigma_l^*)/M_{H^{\pm}}^2$, entering in conflict with constraints from leptonic meson decays.       There is no allowed region when all the observables are considered, though there is if $R(D^*)$ is excluded from the fit.     To explain the current excess in $R(D^*)$ within the framework of 2HDMs one therefore needs a departure from the family universality of the Yukawa couplings, see for example~Refs.~\cite{Fajfer:2012jt,Crivellin:2012ye}.

\begin{table}[tb]
\begin{center}
\caption{\label{tab::SMvsEXP} \it \small SM predictions for the various semileptonic and leptonic decays considered in the analysis, together with their corresponding experimental values. The first uncertainty given always corresponds to the statistical error, and the second, when given, to the theoretical one.}
\vspace{0.2cm}
\doublerulesep 0.8pt \tabcolsep 0.07in
\small{
\begin{tabular}{lccc}
\hline\hline
Observable   					&  SM Prediction					& Exp. Value  \\
\hline \\[-10pt]
$R({D})$ 							& $0.296^{+0.008}_{-0.006}\pm0.015$ & $ 0.438 \pm 0.056$ \\
$R({D^*})$       					& $0.252\pm0.002\pm0.003$ 		& $0.354 \pm 0.026$					 \\
$\Br(B\to \tau \nu_\tau)$ 	&  $(0.79^{+0.06}_{-0.04}\pm0.08)\times 10^{-4}$& $(1.15 \pm 0.23)\times 10^{-4}$	 \\
$\Br(D_s \to \tau \nu_\tau)$	&  $(5.18 \pm 0.08\pm0.17) \times 10^{-2} $& $(5.54 \pm 0.24)\times 10^{-2}$  \\
$\Br(D_s \to \mu \nu)$	& $(5.31 \pm 0.09\pm0.17) \times 10^{-3} $	& $(5.54 \pm 0.24)\times 10^{-3}$ \\
$\Br(D \to \mu \nu)$	&  $(4.11^{+0.06}_{-0.05}\pm0.27) \times 10^{-4} $ & $(3.76 \pm 0.18)\times 10^{-4}$	 \\
$\Gamma(K\to\mu\nu)/\Gamma(\pi\to\mu\nu)$    & $1.333\pm0.004\pm0.026$ & $1.337\pm0.003$ \\
$\Gamma(\tau\to K\nu_\tau)/\Gamma(\tau\to\pi\nu_\tau)$ & $(6.56\pm0.02\pm0.15)\times10^{-2}$ & $(6.46\pm0.10)\times10^{-2}$ \\
\hline\hline
\end{tabular}}
\end{center}
\end{table}

If the observed excess in $R(D^{(*)})$ persists, we would like to gain as much information as possible about the underlying new physics.  Three-body decays like the ones at hand offer considerable information in the differential distributions, see for example Ref.~\cite{Datta:2012qk}.   Interestingly, one can build observables which are not sensitive to charged scalar contributions.  Any deviation from the SM in these observables would indicate unequivocally the presence of non scalar new physics.     One observable of this kind is~\cite{Celis:2012dk}
\begin{equation}
X_{1}(q^2) \equiv R_{D^*}(q^2) - R_{L}^*(q^2)  \,,
\end{equation}
with
\begin{equation}
R_{D^{(*)}}(q^2) = \dfrac{  d\Gamma( \bar B \rightarrow D^{(*)} \tau^- \bar \nu_{\tau}  )/dq^2 }{ d\Gamma( \bar B \rightarrow D^{(*)} \ell^- \bar \nu_{\ell}  )/dq^2} \,, \qquad  R_{L}^*(q^2) = \dfrac{    d\Gamma_{\tau}^{L}/dq^2 }{ d\Gamma_{\ell}^L/dq^2 } \,.
\end{equation}
This observable is not sensitive to charged scalar contributions because a charged Higgs does not contribute to the transverse helicity amplitudes.    Other observables sharing this feature are~\cite{Celis:2012dk}:
\begin{equation}
X_{2}^{D}(q^2) \equiv R_{D}(q^2) \left(   A_{\lambda}^{D}(q^2) +1 \right)  \,, \qquad X_{2}^{D^*}(q^2) \equiv R_{D^*}(q^2) \left(   A_{\lambda}^{D^*}(q^2) +1 \right)\,.
\end{equation}
Here $ A_{\lambda}^{D^{(*)}}(q^2)$ represents the $\tau$-spin asymmetry defined in the center-of-mass frame of the leptonic system,  
\begin{equation}
A_{\lambda}^{D^{(*)}}(q^2)= \dfrac{d\Gamma^{D^{(*)}}[ \lambda_{\tau} =-1/2 ]/dq^2  - d\Gamma^{D^{(*)}}[\lambda_{\tau} =+1/2]/dq^2}{   d\Gamma^{D^{(*)}}[ \lambda_{\tau} =-1/2 ]/dq^2   + d\Gamma^{D^{(*)}}[\lambda_{\tau} =+1/2]/dq^2   } \,.
\end{equation}
CP violating observables which are not sensitive to charged scalar contributions have been defined in Ref.~\cite{Duraisamy:2013pia}.

So far we have discussed constraints coming from flavour processes alone.  Direct and indirect searches for a charged Higgs at colliders place stringent bounds for a light charged Higgs, being complementary to flavour processes.   Precision measurements of the $Z$-width at LEP imply a robust lower bound on the charged Higgs mass $M_{H^{\pm}} > 39.6$~GeV at $95\%$~CL~\cite{Abbiendi:2013hk}.
The limit $M_{H^{\pm}} \gtrsim 80$~GeV was set at LEP from direct charged Higgs searches in the $e^+ e^- \rightarrow H^+ H^-$ channel, assuming that the charged Higgs decays dominantly into fermions~\cite{Abbiendi:2013hk}.   LHC searches for a charged Higgs via top decays $t\rightarrow H^+b$ have been interpreted within the CP-conserving A2HDM in Ref.~\cite{Celis:2013ixa}, putting a limit $| \varsigma_u \varsigma_l |/M_{H^{\pm}}^2 \lesssim 10^{-3}$~GeV$^{-2}$ in the mass range $M_{H^{\pm}} \in [90,150]$~GeV.  

\section{Conclusions}
The BaBar collaboration has observed hints for lepton universality violations in exclusive semileptonic transitions of the type $b \rightarrow c \tau^- \bar \nu_{\tau}$~\cite{Lees:2012xj}.    The present excess in $R(D^*)$ can not be accommodated within the A2HDM taking into account leptonic meson decays in which the charged Higgs also enters at tree level.   None of the 2HDMs with natural flavour conservation can explain the excess in $R(D^{*})$ either, being particular cases of the A2HDM.    If the current excess in $R(D^{(*)})$ persists in the future, the study of differential distributions in these processes will play a crucial role in discriminating between different new physics scenarios.   

\Acknowledgements
I am grateful to the organizers of the conference for the pleasant atmosphere.

\end{document}

%% file: econfmacros.tex
\def\beq{\begin{equation}}
\def\eeq#1{\label{#1}\end{equation}}
\def\eeqn{\end{equation}}

\def\beqa{\begin{eqnarray}}
\def\eeqa#1{\label{#1}\end{eqnarray}}
\def\eeqan{\end{eqnarray}}

\let\bar=\overbar

\def\Dslash{\not{\hbox{\kern-4pt $D$}}}
\def\dslash{\not{\hbox{\kern-2pt $\del$}}}

\def\msb{{\bar{\ssstyle M \kern -1pt S}}}




%% file: CKM_celis.bbl
\begin{thebibliography}{99}

\bibitem{Lees:2012xj}
  J.~P.~Lees {\it et al.}  [BaBar Collaboration],
  Phys.\ Rev.\ Lett.\  {\bf 109} (2012) 101802
  [arXiv:1205.5442 [hep-ex]];
  J.~P.~Lees {\it et al.}  [The BaBar Collaboration],
  Phys.\ Rev.\ D {\bf 88} (2013) 7,  072012
  [arXiv:1303.0571 [hep-ex]].
  

  
\bibitem{Adachi:2009qg}
  I.~Adachi {\it et al.}  [Belle Collaboration],
  arXiv:0910.4301 [hep-ex].
  
  
\bibitem{Bozek:2010xy}
  A.~Bozek {\it et al.}  [Belle Collaboration],
  Phys.\ Rev.\ D {\bf 82} (2010) 072005
  [arXiv:1005.2302 [hep-ex]].
  
  

\bibitem{Pich:2009sp}
  A.~Pich and P.~Tuzon,
  Phys.\ Rev.\ D {\bf 80} (2009) 091702
  [arXiv:0908.1554 [hep-ph]].
  
\bibitem{Jung:2010ik}
  M.~Jung, A.~Pich and P.~Tuzon,
  JHEP {\bf 1011} (2010) 003
  [arXiv:1006.0470 [hep-ph]].



\bibitem{Celis:2012dk}
  A.~Celis, M.~Jung, X.~Q.~Li and A.~Pich,
  JHEP {\bf 1301} (2013) 054
  [arXiv:1210.8443 [hep-ph]];
  A.~Celis, M.~Jung, X.~Q.~Li and A.~Pich,
  J.\ Phys.\ Conf.\ Ser.\  {\bf 447} (2013) 012058
  [arXiv:1302.5992 [hep-ph]].
  
  

\bibitem{Fajfer:2012jt}
  S.~Fajfer, J.~F.~Kamenik, I.~Ni\v sand\v zi\'c and J.~Zupan,
  Phys.\ Rev.\ Lett.\  {\bf 109} (2012) 161801
  [arXiv:1206.1872 [hep-ph]].
  
  
  
\bibitem{Datta:2012qk}
  A.~Datta, M.~Duraisamy and D.~Ghosh,
  Phys.\ Rev.\ D {\bf 86} (2012) 034027
  [arXiv:1206.3760 [hep-ph]].
    
\bibitem{Crivellin:2012ye}
  A.~Crivellin, C.~Greub and A.~Kokulu,
  Phys.\ Rev.\ D {\bf 86} (2012) 054014
  [arXiv:1206.2634 [hep-ph]].
  
\bibitem{He:2012zp}
  X.~G.~He and G.~Valencia,
  Phys.\ Rev.\ D {\bf 87} (2013) 014014
  [arXiv:1211.0348 [hep-ph]].
  
\bibitem{Tanaka:2012nw}
  M.~Tanaka and R.~Watanabe,
  Phys.\ Rev.\ D {\bf 87} (2013) 3,  034028
  [arXiv:1212.1878 [hep-ph]].
  
\bibitem{Ko:2012sv}
  P.~Ko, Y.~Omura and C.~Yu,
  JHEP {\bf 1303} (2013) 151
  [arXiv:1212.4607 [hep-ph]].
  

\bibitem{Biancofiore:2013ki}
  P.~Biancofiore, P.~Colangelo and F.~De Fazio,
  Phys.\ Rev.\ D {\bf 87} (2013) 7,  074010
  [arXiv:1302.1042 [hep-ph]].
  
\bibitem{Dorsner:2013tla}
  I.~Dor\v sner, S.~Fajfer, N.~Ko\v snik and I.~Ni\v sand\v zi\'c,
  JHEP {\bf 1311} (2013) 084
  [arXiv:1306.6493 [hep-ph]].
  
  
  


\bibitem{Sakaki:2013bfa}
  Y.~Sakaki, M.~Tanaka, A.~Tayduganov and R.~Watanabe,
  Phys.\ Rev.\ D {\bf 88} (2013) 9,  094012
  [arXiv:1309.0301 [hep-ph]].

\bibitem{Abada:2013aba}
  A.~Abada, A.~M.~Teixeira, A.~Vicente and C.~Weiland,
  JHEP {\bf 1402} (2014) 091
  [arXiv:1311.2830 [hep-ph]].
  
\bibitem{Duraisamy:2014sna}
  M.~Duraisamy, P.~Sharma and A.~Datta,
  arXiv:1405.3719 [hep-ph].


\bibitem{Abdesselam:2014hkd}
  A.~Abdesselam {\it et al.}  [Belle Collaboration],
  arXiv:1409.5269 [hep-ex].
  
\bibitem{Kamenik:2008tj}
  J.~F.~Kamenik and F.~Mescia,
  Phys.\ Rev.\ D {\bf 78} (2008) 014003
  [arXiv:0802.3790 [hep-ph]];
  S.~Fajfer, J.~F.~Kamenik and I.~Ni\v sand\v zi\'c,
  Phys.\ Rev.\ D {\bf 85} (2012) 094025
  [arXiv:1203.2654 [hep-ph]].
  
  
  
\bibitem{Bailey:2012jg}
  J.~A.~Bailey, A.~Bazavov, C.~Bernard, C.~M.~Bouchard, C.~DeTar, D.~Du, A.~X.~El-Khadra and J.~Foley {\it et al.},
  Phys.\ Rev.\ Lett.\  {\bf 109} (2012) 071802
  [arXiv:1206.4992 [hep-ph]].
  
\bibitem{Becirevic:2012jf}
  D.~Becirevic, N.~Kosnik and A.~Tayduganov,
  Phys.\ Lett.\ B {\bf 716} (2012) 208
  [arXiv:1206.4977 [hep-ph]].
  



    
  
  

  
\bibitem{Duraisamy:2013pia}
  M.~Duraisamy and A.~Datta,
  JHEP {\bf 1309} (2013) 059
  [arXiv:1302.7031 [hep-ph]].

  
  


  
\bibitem{Abbiendi:2013hk}
  G.~Abbiendi {\it et al.}  [ALEPH and DELPHI and L3 and OPAL and LEP Collaborations],
  Eur.\ Phys.\ J.\ C {\bf 73} (2013) 2463
  [arXiv:1301.6065 [hep-ex]].
  
\bibitem{Celis:2013ixa}
  A.~Celis, V.~Ilisie and A.~Pich,
  JHEP {\bf 1312} (2013) 095
  [arXiv:1310.7941 [hep-ph]].
  
  
  
  


\end{thebibliography}
